\documentclass{natureinitial}

\usepackage{lineno}
\usepackage{upgreek}
\usepackage{graphicx}
\usepackage{amsmath,amssymb}

\usepackage{multirow}
\usepackage{booktabs}
\usepackage{bm}
\usepackage{makecell}
\usepackage{color}
\usepackage{verbatim}
\usepackage{hyperref}

\usepackage{xcolor}

\def\HS{$^3_{\Lambda}\hbox{H}$}
\def\HF{$^4_{\Lambda}\hbox{H}$}
\def\aHS{$^3_{\bar{\Lambda}}\overline{\hbox{H}}$}
\def\aHF{\hbox{$^4_{\bar{\Lambda}}\overline{\hbox{H}}$}}

\def\HSTb{$^3_{\Lambda}\hbox{H} {\rightarrow}^3\hbox{He}+\pi^-$}
\def\HFTb{$^4_{\Lambda}\hbox{H} {\rightarrow}^4\hbox{He}+\pi^-$}
\def\aHSTb{$^3_{\bar{\Lambda}}\overline{\hbox{H}} {\rightarrow}^3\overline{\hbox{He}}+\pi^+$}
\def\aHFTb{$^4_{\bar{\Lambda}}\overline{\hbox{H}} {\rightarrow}^4\overline{\hbox{He}}+\pi^+$}

\def\snn200{$\sqrt{s_{NN}} = 200 GeV$}

\title{Observation of the Antimatter Hypernucleus \hbox{$^4_{\bar{\Lambda}}\overline{\hbox{H}}$}}

\author{The STAR Collaboration$*$}

\begin{document}

\maketitle

\begin{abstract}

At the origin of the Universe, asymmetry between the amount of created matter and antimatter led to the matter-dominated Universe as we know today. The origins of this asymmetry remain not completely understood yet. High-energy nuclear collisions create conditions similar to the Universe microseconds after the Big Bang, with comparable amounts of matter and antimatter~\cite{STAR:2005gfr,PHENIX:2004vcz,PHOBOS:2004zne,BRAHMS:2004adc,Muller:2012zq,STAR:2019sjh}.
Much of the created antimatter escapes the rapidly expanding fireball without annihilating, making such collisions an effective experimental tool to create heavy antimatter nuclear objects and study their properties~\cite{STAR:2010gyg,STAR:2011eej,Chen:2018tnh,STAR:2019wjm,ALICE:2015oer,ALICE:2017jmf,ALICE:2015rey,Braun-Munzinger:2018hat}, hoping to shed some light on existing questions on the asymmetry between matter and antimatter. 
Here we report the first observation of the antimatter hypernucleus  \aHF, composed of a $\bar{\Lambda}$ , an antiproton and two antineutrons. The discovery was made through its two-body decay after production in ultrarelativistic heavy-ion collisions by the STAR experiment at the Relativistic Heavy Ion Collider~\cite{ackermann2003star,Harrison:2003sb}. In total, 15.6 candidate \aHF \ antimatter hypernuclei are obtained with an estimated background count of 6.4. The lifetimes of the antihypernuclei \aHS \ and \aHF \ are measured and compared with the lifetimes of their corresponding hypernuclei, testing the symmetry between matter and antimatter. Various production yield ratios among (anti)hypernuclei and (anti)nuclei are also measured and compared with theoretical model predictions, shedding light on their production mechanisms.

\end{abstract}

\begin{figure}[!htb]
    \centering
    \includegraphics[width=0.6\textwidth]{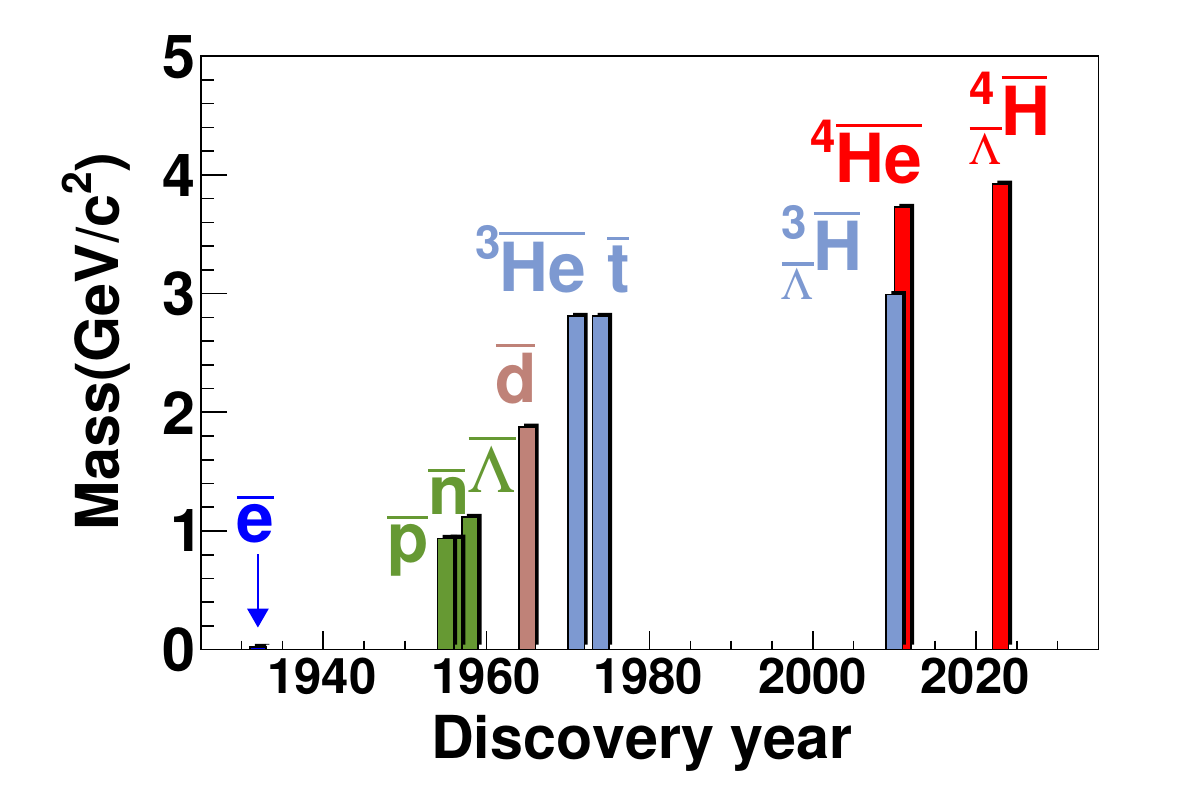}
    \caption{Masses vs. discovery years of selected antimatter particles, including the positron, antinucleons, $\overline{\Lambda}$ and antimatter (hyper)nuclear clusters.}
    \label{fig:fig1}
\end{figure}

In 1928, Paul Dirac found possible solutions with positive and negative energies to his eponymous equation that describes the relativistic quantum behavior of the electron~\cite{Dirac:quantumTheory}. 
It was realized in the following years that the negative-energy solution actually indicates a new particle with the same mass as an electron, but the opposite charge~\cite{Dirac:positron}.
This new particle was discovered by Carl Anderson in cosmic rays in 1932 and named the positron~\cite{Anderson:positron}.
This established the theoretical framework and the experimental foundation for the study of antimatter.
Since then, discovering new, heavier and more complicated antimatter particles and studying their properties have been an important means to explore the nature.
Figure~\ref{fig:fig1} illustrates the masses vs. discovery years of a series of antimatter particles~\cite{Anderson:positron,Chamberlain:antiproton,Cork:antineutron,PhysRevLett.1.179,Massam:antideuteron,Dorfan:antideuteron,Antipov:antiHe3,Vishnevsky:antitriton,STAR:2010gyg,STAR:2011eej}. Among them, \aHF~, whose discovery is described in this paper, is the heaviest antimatter hypernuclear cluster observed to date.

Antimatter readily annihilates with matter, making it difficult to observe antimatter nuclear clusters in the Universe.
However, relativistic heavy-ion collisions can create the quark-gluon-plasma state that existed in the first few microseconds of the Universe after the Big Bang, with nearly equal amounts of matter and antimatter~\cite{STAR:2005gfr,PHENIX:2004vcz,PHOBOS:2004zne,BRAHMS:2004adc,Muller:2012zq,STAR:2019sjh}.
The collision system expands and cools rapidly, allowing some antimatter to decouple from matter. This makes heavy-ion collisions an effective tool to create and study antimatter nuclei or hypernuclei~\cite{Chen:2018tnh,STAR:2019wjm,ALICE:2015oer,ALICE:2017jmf,ALICE:2015rey,Braun-Munzinger:2018hat}.

\renewcommand{\thefootnote}{*}
There are six flavors of quarks, which belong to a group of the most basic building blocks of the visible universe in the standard model of particle physics. 
Among them, the lightest up and down quarks constitute nucleons (i.e., protons and neutrons) in atomic nuclei. 
The strange quark is the third lightest quark. 
Particles with strange quarks tend to decay via the weak interaction, making strange quarks much rarer in nature than the up and down quarks.
A baryon containing at least one strange quark is called a hyperon. For example, the $\Lambda$ hyperon consists of an up, a down and a strange quark.
Like nucleons forming an atomic nucleus, hyperons and nucleons can also constitute a bound state, called a hypernucleus.

In this paper, the Solenoidal Tracker at RHIC (STAR) Collaboration~\cite{ackermann2003star} at the Relativistic Heavy Ion Collider (RHIC)~\cite{Harrison:2003sb} reports the first observation of the antimatter hypernucleus \aHF~, composed of an $\bar{\Lambda}$, an antiproton and two antineutrons \footnote{In \aHF~, H represents the hydrogen with a nuclear charge number $Z$ of 1, 4 is the number of (anti)baryons, and particle symbols with overlines indicate the corresponding antiparticles.}.
We also report the measurements of \HS, \HF, \aHS~ and \aHF~ decay lifetimes, and test matter-antimatter symmetry by hypernucleus-antihypernucleus lifetime comparisons. 
Various production yield ratios among (anti)hypernuclei and (anti)nuclei are measured and compared with theoretical model predictions, shedding light on the production mechanism of (anti)hypernuclei in relativistic heavy-ion collisions.

\section*{(Anti)hypernucleus reconstruction}
RHIC is a gigantic ring-shaped accelerator with a circumference of 3.8 km.
It can accelerate heavy ions (atomic nuclei) to 99.996\% the speed of light.
Pairs of these high-energy heavy ions collide, each producing thousands of final state particles.
The STAR experimental set-up detects and records the produced particles, just like a high-speed 3-dimensional camera.
More than one thousand collisions can be recorded by STAR within a second.
A total of about 6.4 billion U+U, Au+Au, Ru+Ru, and Zr+Zr collision events with center-of-mass energy per colliding nucleon-nucleon pair $\sqrt{s_{NN}}$=193~GeV (U+U) or 200~GeV (other systems) are used in this analysis.

\begin{figure}[htb]
    \centering
    \includegraphics[width=0.9\textwidth]{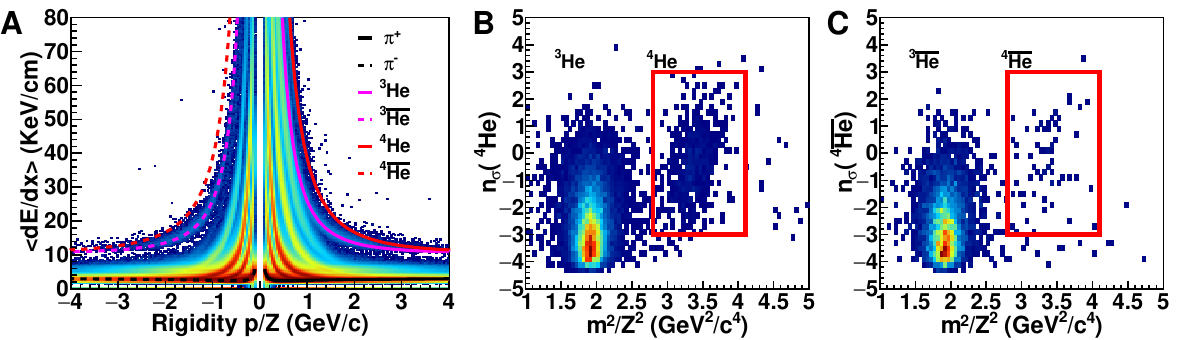}
    \caption{(A) Average energy loss $\langle{dE/dx}\rangle$ versus rigidity of charged particles measured by the TPC. The lines represent the expected trends for $\pi^+$, $^3$He and $^4$He and their corresponding antiparticles. (B) and (C) show $n_{\sigma}(^4{\rm{He}})$ and $n_{\sigma}(^4{\overline{\rm{He}}})$ versus ${m^2}/{Z^2}$. The red boxes indicate the region for $^4\hbox{He}$ and $^4\overline{\hbox{He}}$ candidates.}
    \label{fig:fig2}
\end{figure}

After being created at the collision point, (anti)hypernuclei usually fly only a distance of several centimeters before they decay.
So they can not be seen directly by STAR's main tracking detector, the cylindrical Time Projection Chamber (TPC), which surrounds the collision point with an inner radius of about 60 cm.
Instead, the (anti)hypernucleus is "reconstructed" by tracing back the tracks of its charged daughters to an intersection point where the decay happened.
In this analysis, the two-body decay channels \HSTb, \aHSTb, \HFTb, and \aHFTb are used for (anti)hypernucleus reconstruction.
The charged daughter particles fly out through the TPC, leaving their detectable tracks by loosing energy and ionizing the gas.
The TPC is placed in a 0.5-Tesla solenoidal magnetic field, and the rigidity (momentum over charge) of the charged particle tracks can be measured from their bending in the magnetic field.
Particles with different mass and electrical charge have different average ionization energy loss $\langle{dE/dx}\rangle$ vs. rigidity, as shown in Figure~\ref{fig:fig2}(A), which is used to identify different particles.
Particle identification is further performed with the help of the Time-of-Flight (TOF) detector. 
A particle's squared mass ($m$) over charge ($Z$) ratio, $m^2/Z^2$, is calculated from the rigidity, track length and time of flight.
Figures~\ref{fig:fig2} (B) and (C) show $n_{\sigma}(^4{\rm{He}})$ and  
$n_{\sigma}(^4{\overline{\rm{He}}})$ versus ${m^2}/{Z^2}$, for the selection of $^4\hbox{He}$ and $^4\overline{\hbox{He}}$ candidates. Here $n_{\sigma}$ is the deviation of the measured $\langle{dE/dx}\rangle$ from the expected value for a certain particle species normalized by the resolution $\sigma_{_{dE/dx}}$,
\begin{linenomath*}
\begin{equation}
    n_{\sigma}=\ln\left(\frac{\langle{dE/dx}\rangle}{\langle{dE/dx}\rangle_{\rm th}}\right)/\sigma_{_{dE/dx}}.
\end{equation}
\end{linenomath*}

(Anti)hypernucleus candidates are reconstructed from pairs of selected (anti)helium and $\pi^{\pm}$ tracks.
In order to suppress background from random combinations of particles emitted from the collision point, selections have been applied such that the tracks of the two daughter particles are likely to come from a common decay vertex displaced from the collision point.
The selection cuts on the topological variables are optimized for the best ~\aHS~signal.

\section*{Signals}

\begin{figure}[htb]
    \centering
    \includegraphics[width=0.8\textwidth]{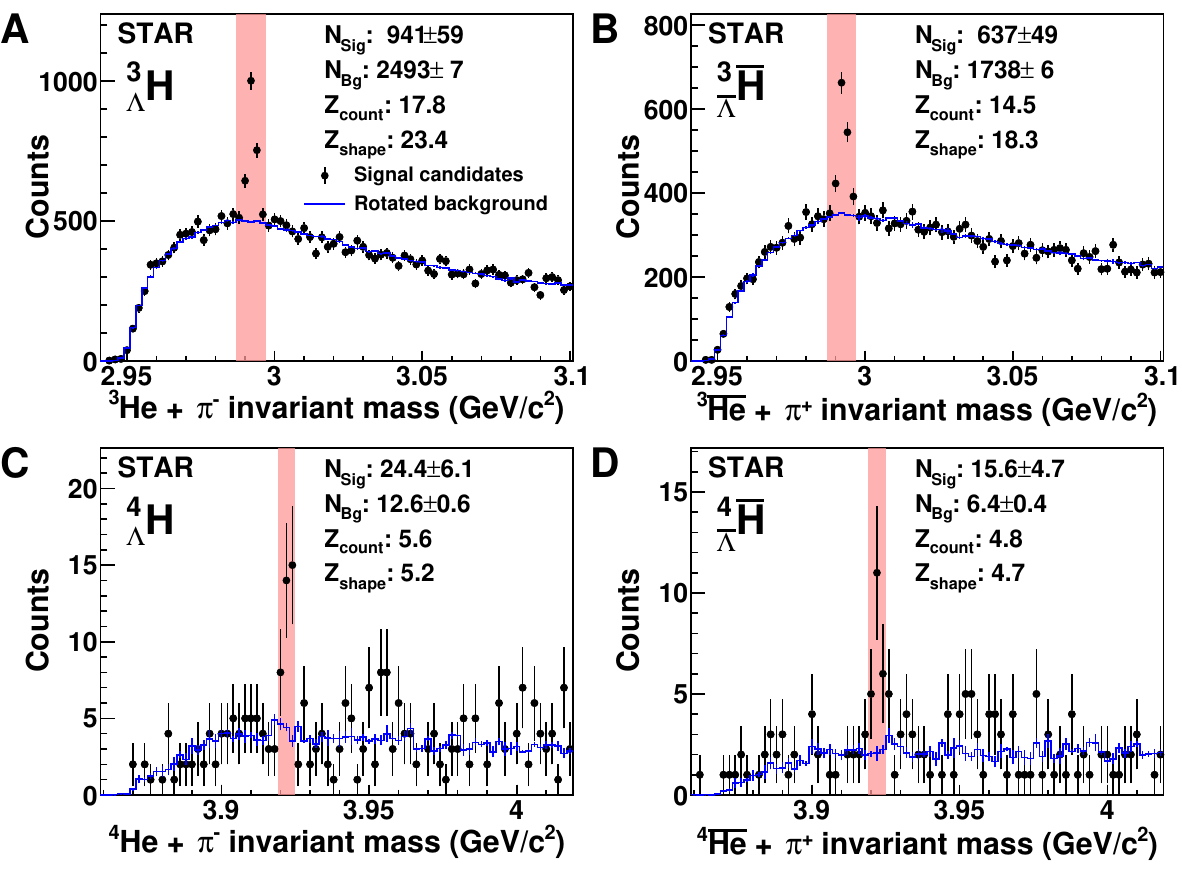}
    \caption{Invariant-mass distributions of 
     $^3\hbox{He}+\pi^-$ (A), $^3\overline{\hbox{He}}+\pi^+$ (B), $^4\hbox{He}+\pi^-$ (C) and $^4\overline{\hbox{He}}+\pi^+$ (D). 
     The solid bands mark the signal invariant-mass regions. The obtained signal count ($N_{\rm Sig}$), background count ($N_{\rm Bg}$), and signal significances ($Z_{\rm count}$ and $Z_{\rm shape}$) are listed in each panel. }
    \label{fig:fig3} 
\end{figure}

To observe the (anti)hypernucleus signals, the invariant mass of their daughter-pair candidates is calculated.  
The invariant mass is the total energy of the daughter particles in their center-of-mass frame, calculated from their 3-dimensional momenta and masses.
According to energy-momentum conservation and Einstein's mass–energy equivalence, the invariant mass of the decay daughters should be equal to the parent-particle mass.
The invariant-mass spectra of reconstructed~\HS,~\aHS,~\HF, and~\aHF~candidates are shown in Fig. \ref{fig:fig3}.
The narrow peaks at the (anti)hypernucleus mass positions are the (anti)hypernucleus decay signals, while the smooth components below are the combinatorial backgrounds.
The combinatorial background invariant-mass distributions are reproduced with a rotation method, in which the (anti)helium nucleus track is randomly rotated around the beam line, so that the decay kinematics of the real signal candidate are destroyed and randomized as the combinatorial background.
The final signal count $N_{\rm Sig}$ is extracted by subtracting the integrated combinatorial background count $N_{\rm Bg}$ from the integral of the signal-candidate distribution in the shaded invariant-mass region in Fig. \ref{fig:fig3}.

In total, $941\pm59$~\HS, $637\pm49$~\aHS, $24.4\pm6.1$~\HF~and $15.6\pm4.7$~\aHF~signal candidates are observed.
The significances are calculated as
\begin{linenomath*}
\begin{equation}
    Z_{\rm count}=\sqrt{2\left[\left( N_{\rm Sig}+N_{\rm Bg} \right)\ln{\left(1+\frac{N_{\rm Sig}}{N_{\rm Bg}}\right)}-N_{\rm Sig} \right]}.
    \label{equation:Zcount} 
\end{equation}
\end{linenomath*}
The significances $Z_{\rm count}$ of ~\HF, and~\aHF~ signals are 5.6 and 4.8 standard deviations ($\sigma$), corresponding to p-values of $1.1\times10^{-8}$ and $7.9\times10^{-7}$, respectively.
The significances are also calculated by comparing the likelihoods of fitting the candidate invariant-mass distributions with a Gaussian-shaped signal plus background with the likelihoods with the hypothesis of pure background. The significances $Z_{\rm shape}$ are obtained as 5.2 and 4.7 $\sigma$ for~\HF~and~\aHF~, respectively.

\section*{Lifetimes and matter-antimatter symmetry test}
Our current knowledge of physics principles suggests  that the initial Universe should have contained equal amounts of matter and antimatter. 
However, the antiproton flux in cosmic rays and other measurements~\cite{AMS:2016oqu} indicate that no large-scale antimatter exists in the vicinity of our galaxy, and the visible universe is almost entirely matter. Naturally, one may ask where the antimatter is, and what causes this matter-antimatter asymmetry in the Universe? 
One expects a matter particle and its corresponding antimatter particle to have the same properties according to the $CPT$ theorem, which states that physical laws should remain unchanged under the combined operation of charge conjugation $C$, parity transformation $P$ and time reversal $T$. 
Comparing the properties like mass and lifetime of a particle and its corresponding antiparticle is an important experimental way to test the $CPT$ symmetry and to search for new mechanisms that cause matter and antimatter asymmetry in the Universe.
Recently, the ALICE and STAR experiments reported that there is no significant mass (binding energy) difference between deuteron and antideuteron~\cite{ALICE:2015rey}, between $^3\hbox{He}$ and $^3\overline{\hbox{He}}$~\cite{ALICE:2015rey} and between \HS~and \aHS~\cite{STAR:2019wjm}.
ALICE has also measured the relative difference between \HS~and \aHS~ lifetimes, which is consistent with zero~\cite{ALICE:2022sco}.

Hypernucleus lifetimes are also an important tool to study the interactions between the hyperons and nucleons within them~\cite{PEREZOBIOL2020135916}, which is a vital nuclear physics input for understanding the inner structure of compact stellar objects like neutron stars~\cite{Gal:2016boi}.
Numerous measurements~\cite{ALICE:2015oer,Prem1964,Keyes1968,Phillips1969,Keyes1970,Keyes1973,Avramenko1992,Rappold2013,STAR2018,STAR2022,Outa1995} show slightly shorter average lifetimes of \HS \ and \HF \ than that of the $\Lambda$ hyperon.
The combined lifetime of \HS \ and \aHS \ has also been measured~\cite{STAR:2010gyg, STAR2018, ALICE2019,ALICE:2022sco}.

\begin{figure}[htbp]
    \centering
    \includegraphics[width=0.9\textwidth]{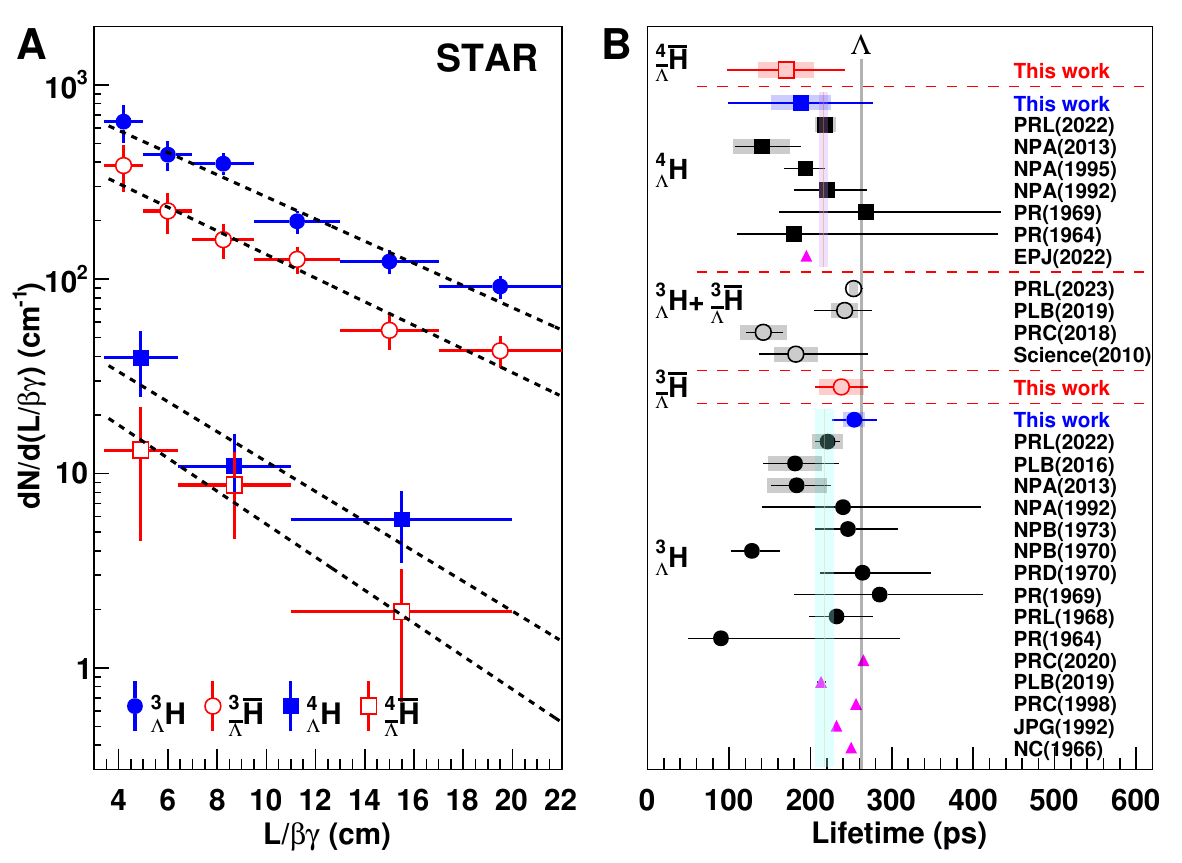}
    \caption{(A) \HS, \aHS, \HF~ and \aHF~ yields versus $L/\beta\gamma$. The vertical error bars represent the statistical uncertainties only.
    (B) Our measured \HS, \aHS, \HF~ and \aHF~ lifetimes compared with world data \cite{STAR:2010gyg,ALICE:2015oer,Prem1964,Keyes1968,Phillips1969,Keyes1970,Bohm:1970se,Keyes1973,Avramenko1992,Rappold2013,STAR2018,ALICE2019,STAR2022,Outa1995,ALICE:2022sco} and theoretical predictions \cite{Rayet1966,Congleton1992,Kamada1998,Gal2019,Hidenbrand2020,Gal2022} (solid triangles). Error bars and boxes show statistical and systematic uncertainties, respectively. Solid vertical lines with shaded regions represent the average lifetimes of \HS \ and \HF \ and their corresponding uncertainties. These are calculated from previous results by a maximum-likelihood fit. The vertical gray line shows the lifetime of the free $\Lambda$.}
    \label{fig:fig4}
\end{figure}

In this study, lifetimes of the (anti)hypernuclei \HS, \HF, \aHS~ and \aHF~ are measured. (Anti)hypernucleus signal yields in $ct = L/\beta\gamma = L/(p/m)$ intervals are obtained as described in the section above, where $c$, $t$, $L$, $\beta$, $\gamma$, $p$ and $m$ represent the speed of light, the decay time in the (anti)hypernucleus rest frame, the measured decay length, the ratio of velocity to $c$, the Lorentz factor of relativistic time dilation, the measured momentum and the (anti)hypernucleus nominal mass, respectively.
The reconstruction efficiencies of \HS, ~\aHS, ~\HF \ and \aHF \ in each $L/\beta\gamma$ bin are evaluated by a Monte Carlo method in which (anti)hypernuclei are simulated using the GEANT3 software package and embedded in real collision events. In this way, the simulated (anti)hypernuclei are reconstructed in a realistic environment.
Efficiency-corrected yields of \HS, ~\aHS, ~\HF \ and \aHF \ as a function of $L/\beta\gamma$ are shown in Fig.\ref{fig:fig4}(A). The lifetimes $\tau$ are extracted by fitting the data with the exponential decay law $N(t) = N_{0}\exp(-t/\tau) = N_{0}\exp(-(L/\beta\gamma)/c\tau)$.

The extracted lifetimes are
\begin{eqnarray}
\tau\left(^{3}_{\Lambda}\hbox{H} \right) &=& 254\pm 28(\rm{stat.})\pm 14(\rm{sys.}) \rm{ps}	 \nonumber ,\\
\tau\left(^{3}_{\bar{\Lambda}}\overline{\hbox{H}} \right) &=& 238\pm 33(\rm{stat.})\pm 28(\rm{sys.}) \rm{ps}	 \nonumber ,\\
\tau\left(^{4}_{\Lambda}\hbox{H} \right) &=& 188\pm 89(\rm{stat.})\pm 37(\rm{sys.}) \rm{ps}	 \nonumber ,\\
\tau\left(^{4}_{\bar{\Lambda}}\overline{\hbox{H}} \right) &=& 170\pm 72(\rm{stat.})\pm 34(\rm{sys.}) \rm{ps}  \nonumber .
\end{eqnarray}
As shown in Fig.~\ref{fig:fig4}(B), our results are consistent with most existing measurements within uncertainties, \cite{STAR:2010gyg,ALICE:2015oer,Prem1964,Keyes1968,Phillips1969,Keyes1970,Keyes1973,Avramenko1992,Rappold2013,STAR2018,ALICE2019,STAR2022,Outa1995,ALICE:2022sco} and theory predictions \cite{Rayet1966,Congleton1992,Kamada1998,Gal2019,Hidenbrand2020,Gal2022}.
The lifetime differences between hypernuclei and their corresponding antihypernuclei are $\tau\left({^3_{\Lambda}\hbox{H}}\right)-\tau\left({^3_{\bar{\Lambda}}\overline{\hbox{H}}}\right)=$[16 $\pm$ 43(stat.) $\pm$ 20(sys.)] ps and ~$\tau\left({^4_{\Lambda}\hbox{H}}\right)-\tau\left({^4_{\bar{\Lambda}}\overline{\hbox{H}}}\right)=$[18 $\pm$ 115(stat.) $\pm$ 46(sys.)] ps. 
Both are consistent with zero within uncertainties, showing no difference between the properties of matter particles and those of their corresponding antimatter particles. This is a new test of the $CPT$ symmetry.

\section*{Yield ratios}
The (anti)nucleus and (anti)hypernucleus production yields carry information about their production mechanism in relativistic heavy-ion collisions.
Collisions at RHIC energies create fireballs with a temperature of several hundred MeV \cite{andronic2011production}, that corresponds to 10$^{12}$ K, while the (anti)nuclei and (anti)hypernuclei have typical binding energies of merely several MeV per (anti)baryon.
Thus, it is often imagined that these fragile objects are produced in the last stage of the collision-system evolution, via coalescence of (anti)hyperons and (anti)nucleons that are by chance close in both coordinate and momentum space~\cite{sato1981coalescence,steinheimer2012hypernuclei,Sun:2015ulc}. 
As observed in earlier measurements~\cite{STAR:2011eej,ALICE:2017jmf}, the probability to coalesce decreases by 2-3 orders of magnitude with each additional (anti)baryon.
Since the $\Lambda$ baryon is heavier than the nucleons, it takes more energy to be created. There are fewer $\Lambda$ baryons than protons and neutrons created in the fireballs, thus (anti)hypernucleus production yields are usually lower than those of (anti)nuclei with the same baryon numbers~\cite{STAR:2010gyg,ALICE:2015oer}.
These baryon number and strangeness dependencies of particle production yields can also be well described by the statistical thermal model~\cite{andronic2011production}, which assumes all particles to be in a thermal and chemical equilibrium.
The parameters of the statistical thermal model (chemical freeze-out temperature $T$ and baryon chemical potential $\mu_{B}$) can be obtained by a simultaneous fit to all existing measured particle yields.

This analysis uses a combination of data from U+U, Au+Au, Ru+Ru and Zr+Zr collision systems, with different particle production yields. 
Thus the absolute (anti)hypernuclear production yields in this mixture of collision systems are not well-defined physics quantities to measure.
Instead, we measure various yield ratios among (anti)nuclei and (anti)hypernuclei with the same number of (anti)baryons.
In this way, the yield differences due to different collision-system sizes will largely cancel out.
The measurement is done with particles in a phase-space region of rapidity $|y|<0.7$ (i.e., the velocity component along the beam direction in the range of $|v_{z}|<0.604c$) and $0.7<p_{T}/m<1.5$, where $p_{T}$ is the momentum in the plane transverse to the beam direction.
Detector acceptance, efficiency and decay branching fractions are corrected for.
Due to the lack of conclusive theoretical or experimental results, we assume 0.25 as the decay branching fraction of \HSTb and \aHSTb~\cite{STAR:2010gyg,ALICE:2015oer,STAR2022,E864:2002xhb}, and 0.5 for \HFTb and \aHFTb~\cite{STAR2022,E864:2002xhb}. 
$^3\hbox{He}$, $^3\overline{\hbox{He}}$, $^4\hbox{He}$, and $^4\overline{\hbox{He}}$ yields are corrected for contributions from \HS, \aHS, \HF, and \aHF \ decays when calculating the ratios.

\begin{figure}[htbp]
    \centering
    \includegraphics[width=0.8\textwidth]{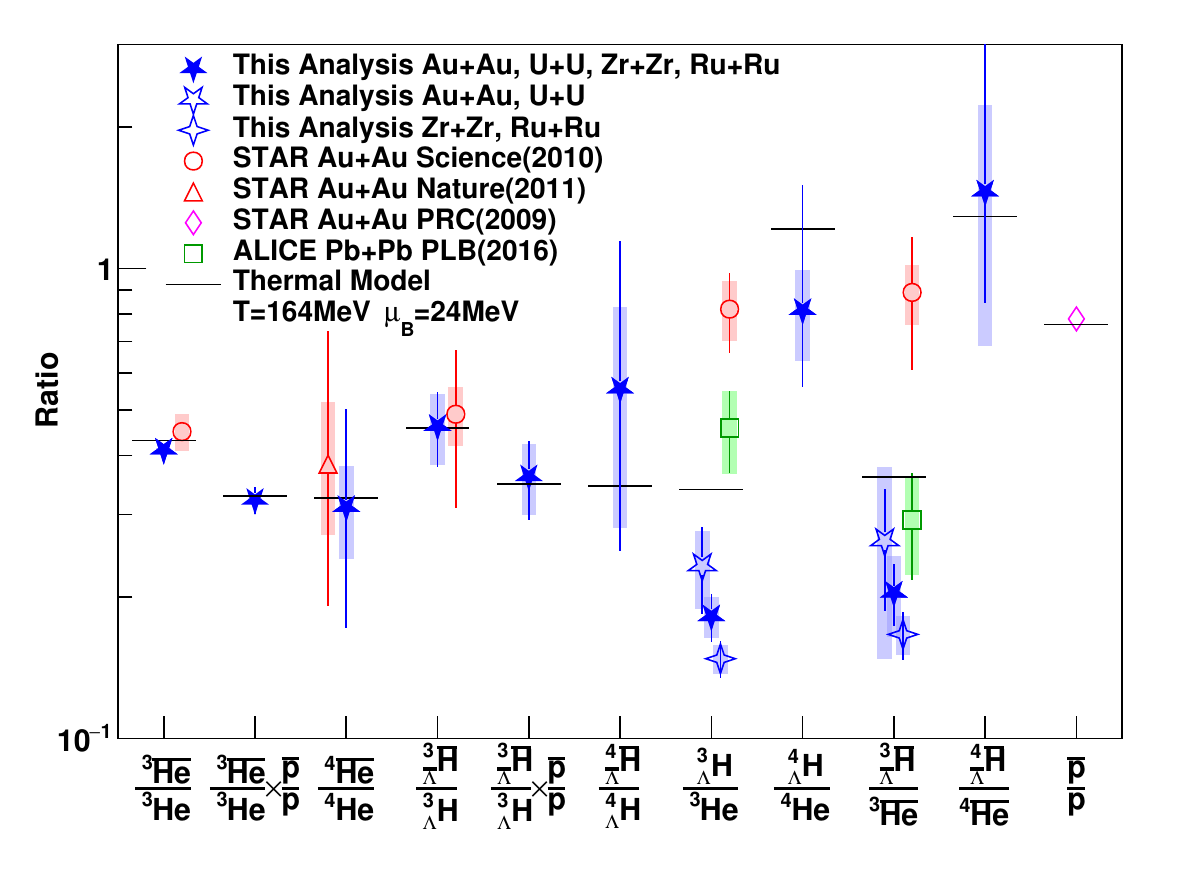}
    \caption{Production-yield ratios among the various (anti)nuclei and (anti)hypernuclei with the same number of (anti)baryons. Results combining all collision systems in this work are shown by filled stars. Open stars show results with only U+U and Au+Au collisions, while quadrangular stars show results with only Zr+Zr and Ru+Ru collisions. Statistical uncertainties and systematic uncertainties are shown by vertical bars and shaded boxes, respectively.
    The decay branching fraction of \HSTb and \aHSTb is assumed to be 0.25~\cite{Kamada1998,Gloeckle:1998ty}, and the branching fraction of \HFTb and \aHFTb is assumed to be 0.5~\cite{Kumagai-Fuse:1995arx,Outa:1998tt}. 
    Previous measurement results~\cite{STAR:2010gyg,STAR:2011eej,ALICE:2015oer,STAR:2006uve} and statistical-thermal-model predictions~\cite{andronic2011production} are also shown for comparison. }
    \label{fig:fig5}
\end{figure}

Figure~\ref{fig:fig5} shows the measured particle production yield ratios and a comparison to previous experimental results\cite{STAR:2010gyg,STAR:2011eej,ALICE:2015oer,STAR:2006uve}, as well as the statistical thermal model predictions\cite{andronic2011production}.
Since the $^3_{\Lambda}\hbox{H}/^3\hbox{He}$ and $^3_{\bar{\Lambda}}\overline{\hbox{H}}/^3\overline{\hbox{He}}$ ratios are expected to increase with the collision-system size\cite{Sun:2018mqq,Bellini:2018epz}, we have also measured them in large (U+U, Au+Au) and small (Zr+Zr, Ru+Ru) systems separately, in order to compare with existing measurements.
The measured particle ratios are consistent with previous measurements, while we note that the $^3_{\Lambda}\hbox{H}/^3\hbox{He}$ and $^3_{\bar{\Lambda}}\overline{\hbox{H}}/^3\overline{\hbox{He}}$ ratios in U+U and Au+Au collisions are lower than previous STAR results~\cite{STAR:2010gyg} by 2.8 and 1.9 $\sigma$, respectively.

Various antimatter-over-matter particle-yield ratios are measured to be below unity because the colliding heavy ions carry positive baryon numbers, and consequently the collision system has positive baryon chemical potential.
We also observe that $^4\overline{\hbox{He}}/^4\hbox{He}$ $\sim$ $^3\overline{\hbox{He}}/^3\hbox{He}\times\overline{\hbox{p}}/\hbox{p}$, $^4_{\bar{\Lambda}}\overline{\hbox{H}}/^4_{\Lambda}\hbox{H}$ $\sim$ $^3_{\bar{\Lambda}}\overline{\hbox{H}}/^3_{\Lambda}\hbox{H}\times\overline{\hbox{p}}/\hbox{p}$, $^4_{\Lambda}\hbox{H}/^4\hbox{He}$ $\sim$ $4\times{^3_{\Lambda}\hbox{H}/^3\hbox{He}}$, and $^4_{\bar{\Lambda}}\overline{\hbox{H}}/^4\overline{\hbox{He}}$ $\sim$ $4\times{^3_{\bar{\Lambda}}\overline{\hbox{H}}}/^3\overline{\hbox{He}}$, as expected in the coalescence~\cite{sato1981coalescence,steinheimer2012hypernuclei} picture of (anti)nucleus and (anti)hypernucleus production.
Here the factors 4 are introduced because both spin-0 and spin-1 states of \HF~ have enough binding energy so that no energetically allowed strong decay channels exist for them. So the spin-1 state, with a spin degeneracy of 3, will decay electromagnetically to the spin-0 ground state. This enhances the total measured \HF~ and \aHF~ production yield by a factor of 4, compared to $^4\hbox{He}$ and $^4\overline{\hbox{He}}$ which have only a spin-0 state~\cite{STAR2022}.
Considering this spin-degeneracy effect, the statistical-thermal-model~\cite{andronic2011production} predictions also match our measurements, except that the measured $^3_{\Lambda}\hbox{H}/^3\hbox{He}$ ratio is slightly lower than the statistical-thermal-model prediction.
This difference, which is currently not statistically significant, may be explained by the very small binding energy of \HS, which implies that the spatial extent of the \HS~ wave function is comparable to the whole collision system ~\cite{E864:1999zqh,Sun:2018mqq,Bellini:2018epz}.

In general, our measured particle yield ratios are consistent with the expectation of the coalescence picture of (anti)nucleus and (anti)hypernucleus production and the statistical thermal model.
Despite an enhancement factor of 4 due to the spin-degeneracy effect, the $^4_{\bar{\Lambda}}\overline{\hbox{H}}$ production yield is still about 2 orders of magnitude lower than that of $^3_{\bar{\Lambda}}\overline{\hbox{H}}$~\cite{Sun:2015ulc}. 
Fourteen years after the discovery of the first antihypernucleus \aHS, 15.6 $^4_{\bar{\Lambda}}\overline{\hbox{H}}$ signal candidates are reconstructed and identified out of 6.4 billion collision events in this study, which is a significant step forward in the experimental research of antimatter.


\section*{Acknowledgement}
We thank the RHIC Operations Group and RCF at BNL, the NERSC Center at LBNL, and the Open Science Grid consortium for providing resources and support. This work was supported in part by the Office of Nuclear Physics within the U.S. DOE Office of Science, the U.S. National Science Foundation, National Natural Science Foundation of China, Chinese Academy of Science, the Ministry of Science and Technology of China and the Chinese Ministry of Education, the Higher Education Sprout Project by Ministry of Education at NCKU, the National Research Foundation of Korea, Czech Science Foundation and Ministry of Education, Youth and Sports of the Czech Republic, Hungarian National Research, Development and Innovation Office, New National Excellency Programme of the Hungarian Ministry of Human Capacities, Department of Atomic Energy and Department of Science and Technology of the Government of India, the National Science Centre and WUT ID-UB of Poland, the Ministry of Science, Education and Sports of the Republic of Croatia, German Bundesministerium f\"ur Bildung, Wissenschaft, Forschung and Technologie (BMBF), Helmholtz Association, Ministry of Education, Culture, Sports, Science, and Technology (MEXT) and Japan Society for the Promotion of Science (JSPS). We thank the Joint Department for Nuclear Physics, co-founded by the Lanzhou University and Institute of Modern Physics, Chinese Academy of Sciences, for the contributions of its students Junlin Wu and Fengyi Zhao to this paper.

\section*{Data availability}
All raw data for this study were collected using the STAR detector at Brookhaven National Laboratory and are not available to the public. Derived data supporting the findings of this study are publicly available in the HEPData repository~\href{https://www.hepdata.net/record/145132}{https://www.hepdata.net/record/145132} or from the corresponding author on request.

\section*{Code availability}
The codes to process raw data collected by the STAR detector are publicly available at \\\href{https://github.com/star-bnl}{https://github.com/star-bnl}. The codes to analyse the produced data are not publicly available.

\section*{Contributions, inclusion and ethics}
All authors contributed extensively and do not have any competing interests.

\section*{Reference}
\bibliographystyle{naturemag}


\section*{Author:}
\author{
M.~I.~Abdulhamid$^{4}$,
B.~E.~Aboona$^{56}$,
J.~Adam$^{16}$,
L.~Adamczyk$^{2}$,
J.~R.~Adams$^{41}$,
I.~Aggarwal$^{43}$,
M.~M.~Aggarwal$^{43}$,
Z.~Ahammed$^{63}$,
E.~C.~Aschenauer$^{6}$,
S.~Aslam$^{28}$
% ...
}

\normalsize{\rm{(STAR Collaboration)}}

\section*{Affiliations:}
\normalsize{$^{1}$Abilene Christian University, Abilene, Texas   79699}\newline
\normalsize{$^{2}$AGH University of Krakow, FPACS, Cracow 30-059, Poland}\newline
\normalsize{$^{3}$Argonne National Laboratory, Argonne, Illinois 60439}\newline
\normalsize{$^{4}$American University in Cairo, New Cairo 11835, Egypt}\newline
\normalsize{$^{5}$Ball State University, Muncie, Indiana, 47306}\newline
\normalsize{$^{6}$Brookhaven National Laboratory, Upton, New York 11973}\newline
\normalsize{$^{7}$University of Calabria \& INFN-Cosenza, Rende 87036, Italy}\newline
\normalsize{$^{8}$University of California, Berkeley, California 94720}\newline
\normalsize{$^{9}$University of California, Davis, California 95616}\newline
\normalsize{$^{10}$University of California, Los Angeles, California 90095}\newline
\normalsize{$^{11}$University of California, Riverside, California 92521}\newline
\normalsize{$^{12}$Central China Normal University, Wuhan, Hubei 430079 }\newline
\normalsize{$^{13}$University of Illinois at Chicago, Chicago, Illinois 60607}\newline
\normalsize{$^{14}$Chongqing University, Chongqing, 401331}\newline
\normalsize{$^{15}$Creighton University, Omaha, Nebraska 68178}\newline
\normalsize{$^{16}$Czech Technical University in Prague, FNSPE, Prague 115 19, Czech Republic}\newline
\normalsize{$^{17}$Technische Universit\"at Darmstadt, Darmstadt 64289, Germany}\newline
\normalsize{$^{18}$National Institute of Technology Durgapur, Durgapur - 713209, India}\newline
\normalsize{$^{19}$ELTE E\"otv\"os Lor\'and University, Budapest, Hungary H-1117}\newline
\normalsize{$^{20}$Frankfurt Institute for Advanced Studies FIAS, Frankfurt 60438, Germany}\newline
\normalsize{$^{21}$Fudan University, Shanghai, 200433 }\newline
\normalsize{$^{22}$Guangxi Normal University, Guilin, 541004}\newline
\normalsize{$^{23}$University of Heidelberg, Heidelberg 69120, Germany }\newline
\normalsize{$^{24}$University of Houston, Houston, Texas 77204}\newline
\normalsize{$^{25}$Huzhou University, Huzhou, Zhejiang  313000}\newline
\normalsize{$^{26}$Indian Institute of Science Education and Research (IISER), Berhampur 760010 , India}\newline
\normalsize{$^{27}$Indian Institute of Science Education and Research (IISER) Tirupati, Tirupati 517507, India}\newline
\normalsize{$^{28}$Indian Institute Technology, Patna, Bihar 801106, India}\newline
\normalsize{$^{29}$Indiana University, Bloomington, Indiana 47408}\newline
\normalsize{$^{30}$Institute of Modern Physics, Chinese Academy of Sciences, Lanzhou, Gansu 730000 }\newline
\normalsize{$^{31}$University of Jammu, Jammu 180001, India}\newline
\normalsize{$^{32}$Kent State University, Kent, Ohio 44242}\newline
\normalsize{$^{33}$University of Kentucky, Lexington, Kentucky 40506-0055}\newline
\normalsize{$^{34}$Lawrence Berkeley National Laboratory, Berkeley, California 94720}\newline
\normalsize{$^{35}$Lehigh University, Bethlehem, Pennsylvania 18015}\newline
\normalsize{$^{36}$Max-Planck-Institut f\"ur Physik, Munich 80805, Germany}\newline
\normalsize{$^{37}$Michigan State University, East Lansing, Michigan 48824}\newline
\normalsize{$^{38}$National Institute of Science Education and Research, HBNI, Jatni 752050, India}\newline
\normalsize{$^{39}$National Cheng Kung University, Tainan 70101 }\newline
\normalsize{$^{40}$Nuclear Physics Institute of the CAS, Rez 250 68, Czech Republic}\newline
\normalsize{$^{41}$The Ohio State University, Columbus, Ohio 43210}\newline
\normalsize{$^{42}$Institute of Nuclear Physics PAN, Cracow 31-342, Poland}\newline
\normalsize{$^{43}$Panjab University, Chandigarh 160014, India}\newline
\normalsize{$^{44}$Purdue University, West Lafayette, Indiana 47907}\newline
\normalsize{$^{45}$Rice University, Houston, Texas 77251}\newline
\normalsize{$^{46}$Rutgers University, Piscataway, New Jersey 08854}\newline
\normalsize{$^{47}$University of Science and Technology of China, Hefei, Anhui 230026}\newline
\normalsize{$^{48}$South China Normal University, Guangzhou, Guangdong 510631}\newline
\normalsize{$^{49}$Sejong University, Seoul, 05006, South Korea}\newline
\normalsize{$^{50}$Shandong University, Qingdao, Shandong 266237}\newline
\normalsize{$^{51}$Shanghai Institute of Applied Physics, Chinese Academy of Sciences, Shanghai 201800}\newline
\normalsize{$^{52}$Southern Connecticut State University, New Haven, Connecticut 06515}\newline
\normalsize{$^{53}$State University of New York, Stony Brook, New York 11794}\newline
\normalsize{$^{54}$Instituto de Alta Investigaci\'on, Universidad de Tarapac\'a, Arica 1000000, Chile}\newline
\normalsize{$^{55}$Temple University, Philadelphia, Pennsylvania 19122}\newline
\normalsize{$^{56}$Texas A\&M University, College Station, Texas 77843}\newline
\normalsize{$^{57}$University of Texas, Austin, Texas 78712}\newline
\normalsize{$^{58}$Tsinghua University, Beijing 100084}\newline
\normalsize{$^{59}$University of Tsukuba, Tsukuba, Ibaraki 305-8571, Japan}\newline
\normalsize{$^{60}$University of Chinese Academy of Sciences, Beijing, 101408}\newline
\normalsize{$^{61}$United States Naval Academy, Annapolis, Maryland 21402}\newline
\normalsize{$^{62}$Valparaiso University, Valparaiso, Indiana 46383}\newline
\normalsize{$^{63}$Variable Energy Cyclotron Centre, Kolkata 700064, India}\newline
\normalsize{$^{64}$Warsaw University of Technology, Warsaw 00-661, Poland}\newline
\normalsize{$^{65}$Wayne State University, Detroit, Michigan 48201}\newline
\normalsize{$^{66}$Wuhan University of Science and Technology, Wuhan, Hubei 430065}\newline
\normalsize{$^{67}$Yale University, New Haven, Connecticut 06520}\newline
\normalsize{{$^{*}${\rm Deceased}}}

\end{document}


\setcounter{page}{29}

\begin{methods}

\subsection{Event Sample and Trigger Selection.} 
This analysis used 606 million and 624 million $\sqrt{s_{NN}}=200$ GeV Au+Au collision events obtained in years 2010 and 2011, 512 million $\sqrt{s_{NN}}=193$ GeV U+U collision events from year 2012, and 4.7 billion $\sqrt{s_{NN}}=200$ GeV Ru+Ru and Zr+Zr collision events from year 2018.

The majority of events were collected with minimum-bias (MB) triggers. The MB trigger is designed to accept the events with different impact parameters as equally as possible.  The MB triggers required a coincidence between either the vertex-position detectors (VPD) or the zero-degree calorimeters (ZDC).  The VPD~\cite{StarVPD} is a pair of timing detectors mounted directly around the beampipe that cover approximately half of the phase space over the pseudorapidity region $4.2 < |\eta| < 5.2$.  The ZDC~\cite{StarZDC} is a pair of hadronic calorimeters located at $|\eta| > 6.6$ that detect spectator neutrons emerging from the heavy-ion collisions.

Often the MB triggers were highly prescaled to reserve a fraction of the data-acquisition bandwidth for triggers on rare processes.  Events that satisfied ``central'' or ``non-photonic electron'' triggers were included in the analysis to enhance the overall statistics. 
The central triggers combined multiplicity information from the time-of-flight system~\cite{llope2012multigap} with spectator-neutron multiplicity information from the ZDCs to select collisions with small impact parameters.
The non-photonic electron triggers, intended primarily to select events containing electrons from charm- and bottom-quark decays, required a large transverse energy deposition ($E_T > 2.6$, 3.5, or 4.2 GeV) in at least one $\Delta\eta \times \Delta\phi = 0.05 \times 0.05$ tower in the barrel electromagnetic calorimeter~\cite{StarEMC}. 
They have a high probability to trigger on events containing antinuclei, which may annihilate in the electromagnetic calorimeter. Events triggered by the ``central'' or ``non-photonic electron'' triggers were not used in the yield ratios analysis to avoid potential biases.

The reconstructed collision point, called the primary vertex, is required to be within 2~cm of the beam line and within 40~cm along the beam line from the detector center.

\subsection{Daughter-Particle Identification.} 
Information from the TPC~\cite{anderson2003star} and the TOF~\cite{llope2012multigap} are combined for particle identification.
The cylindrical TPC has full azimuthal coverage in the pseudorapidity range $-1<\eta<1$.
In order to ensure good track quality, a minimum of 20 measured points in TPC is required for all tracks used in this analysis.
The average particle energy loss $\langle{dE/dx}\rangle$ vs. rigidity is theoretically described with the Bichsel function~\cite{bichsel2006method}.
A selected $^{3}\hbox{He}$ or $^{3}\overline{\hbox{He}}$ candidate should satisfy $|n_{\sigma^{3}\rm{He}}|<3$. If the track has matched TOF hit information, it should also satisfy the condition $1.0<{m^2}/{Z^2}<3.0$ (GeV$/c^2)^2$.
For $^{4}\hbox{He}$ and $^{4}\overline{\hbox{He}}$ selection, in addition to 
$|n_{\sigma^{4}\rm{He}}|<3$, it is also required that $2.8<{m^2}/{Z^2}<4.1$ (GeV/$c^2)^2$ if a matching TOF hit is present or $|n_{\sigma^3\rm{He}}|>3.5$ if there is no TOF match, in order to minimize contamination from $^{3}\hbox{He}$ and $^{3}\overline{\hbox{He}}$, which have much higher production yields.
In order to reject background $^{3}\hbox{He}$ and $^{4}\hbox{He}$ knocked out from the beam pipe and other materials, the distance-of-closest approach ($DCA$) between the $^{3}\hbox{He}$ or $^{4}\hbox{He}$ trajectory and the primary vertex is required to be within 1 cm. 
This $DCA$ requirement is not applied to $^{3}\overline{\hbox{He}}$ and $^{4}\overline{\hbox{He}}$ since there are no knock-out antinuclei. 
The daughter $\pi^{\pm}$ from (anti)hypernucleus decay is identified by requiring $|n_{\sigma\pi^{\pm}}| < 3$. A $m^2/Z^2$ cut is also applied if the track is associated with a TOF hit.

\subsection{Topological Reconstruction.} 
(Anti)hypernucleus candidates are reconstructed from the selected $\pi^{\pm}$ and (anti)helium nucleus tracks by the Kalman-Filter (KF) Particle Finder package~\cite{Kisel:2012xia,Maksym2016,Ju:2023xvg,Ablyazimov:2015pka}, which is based on the Kalman filter method.
The decay topology of a hypernucleus, as illustrated in Fig.~\ref{fig:TopoGeom}, is characterized by several variables: ${\chi}^2_{topo}$ describing the deviation of the reconstructed mother particle's path from the primary vertex, ${\chi}^2_{NDF}$ describing the deviation between the two daughter tracks at the decay vertex, ${\chi}^2_{primary}$ describing the deviation of the decay-daughter track from the primary vertex, the decay length ($L$), and $L$ over its uncertainty ($L/dL$).
The selection cuts on these topological variables are optimized for the best ~\ahtl~signal, instead of ~\ahfl~signal, in order to avoid any bias towards a better signal and a larger yield of ~\ahfl~due to statistical fluctuations.
This bias due to fluctuations is much smaller for ~\ahtl~ because of its large signal significance. 
The optimized topological-selection cuts are listed in Tab.~\ref{cut}.
Most selections are applied such that the two daughter tracks are likely to come from a common decay vertex with significant displacement from the collision point.
Since the (anti)helium is much heavier than the decay daughter pion, the momentum and track direction of the (anti)helium are very close to those of the parent (anti)hypernuclei.
Thus the (anti)helium $DCA$ due to decay is too small to be clearly observed with STAR-TPC tracking resolution, and a lower limit on He $\chi^2_{primary}$ does not help to improve the signal.
The very loose upper limit on He $\chi^2_{primary}$ is used here to reject background helium candidate tracks that are too far away from the collision point, for example, from pile-up events.

\begin{figure}[htb]
    \centering
    \includegraphics[width=0.8\textwidth]{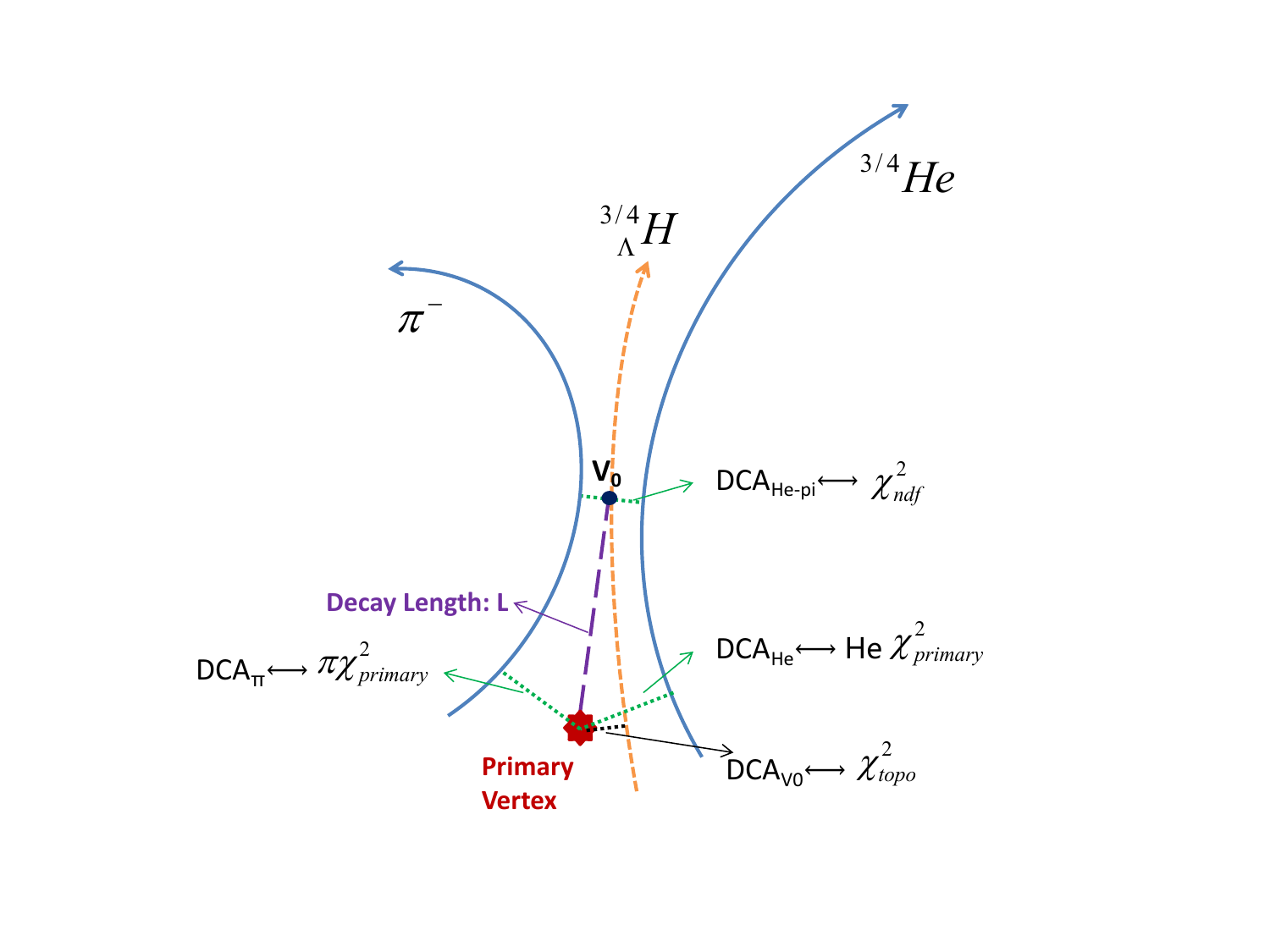}
    \caption{Illustration of the decay topology of a hypernucleus and the variables for the selection criteria.}
    \label{fig:TopoGeom}
\end{figure}

\begin{table}[htbp]
    \centering
	\caption{Topological cuts for (anti)hypernucleus selection.}
	\label{cut}
	\begin{tabular}{cccccccc}
			\hline
			Particles & $\chi^2_{topo}$ & $ \chi^2_{NDF} $ & $\pi$ $\chi^2_{primary} $ & He $ \chi^2_{primary} $ & $L$(cm) & $L/dL$\\
			\hline
            \htl, \hfl   & $<2$ & $<5$  & $>10$ & $<2000$ & $>3.5$ & $>3.4$\\
            \ahtl, \ahfl & $<3$ & $<5$  & $>10$ & $<2000$ & $>3.5$ & $>3.4$\\
            \hline
	\end{tabular}
\end{table}

The invariant mass of a (anti)hypernucleus candidate is calculated as 
$\sqrt{(E_{He}+E_{\pi})^{2}-(\mathbf{p}_{He}+\mathbf{p}_{\pi})^{2}}$, where $E$ and $\mathbf{p}$ are the energy and 3-dimensional momentum of the daughter particles.

\subsection{Background Subtraction.}
The invariant-mass distributions of the combinatorial backgrounds are reproduced with the rotation method. 
Before a helium track is paired with a pion track, its azimuthal angle $\phi$ is rotated randomly in a range of $[30^{\circ},330^{\circ}]$. 
This procedure is repeated 50 times to increase the statistics. Then the same topological-selection cuts as for signal-candidate selection are applied for the rotational background.
They are then scaled so that the their integrals in two side-band regions ($2.941\sim2.987$ GeV/$c^2$ and $2.997\sim3.101$ GeV/$c^2$ for \htl~ and \ahtl, $3.859\sim3.919$ GeV/$c^2$ and $3.925\sim4.019$ GeV/$c^2$ for \hfl~ and \ahfl) are equal to the integrals of the signal-candidate invariant-mass distributions in the same regions. The statistical uncertainties in the rotational background are obtained with a bootstrapping method.
After that, the signal counts are extracted by subtracting the integrals of the scaled combinatorial-background distributions from the integrals of the signal-candidate distributions in the signal invariant-mass regions ($2.987\sim2.997$ GeV/$c^2$ for \htl~ and \ahtl, $3.919\sim3.925$ GeV/$c^2$ for \hfl~ and \ahfl)~\cite{HypernuclearDataBase}.

\subsection{Significance Calculation.}
The signal significances in this analysis are obtained by calculating the likelihood ratios between the hypothesis of pure background and that of signal plus background. This is conducted both by counting the signal and background in a predefined signal invariant-mass region, and by fitting the candidate invariant-mass distribution without and with the signal. In the counting method, the significance is calculated by the asymptotic formula~\cite{Cowan:2010js}
\begin{linenomath*}
\begin{equation}
    Z_{\rm count}=\sqrt{2\left[\left( N_{\rm Sig}+N_{\rm Bg} \right)\ln{\left(1+\frac{N_{\rm Sig}}{N_{\rm Bg}}\right)}-N_{\rm Sig} \right]},
    \label{equation:Zcount} 
\end{equation}
\end{linenomath*}
where the signal count $N_{\rm Sig}$ and background count $N_{\rm Bg}$ are obtained as described in the previous paragraph. In the fitting method, the candidate invariant-mass distribution is firstly fit by pure rotational background with a free scaling factor, then fit by rotational background plus a Gaussian-shaped signal. The Gaussian shape is due to the measured daughter-particle momentum resolution, which is propagated to the calculated invariant mass. All the Gaussian parameters are free in the fit. The likelihood ratios between the fits without and with the Gaussian-shaped signals are used to calculate the significances $Z_{\rm shape}$.

\subsection{Efficiency Correction.}
A correction is applied for the detector acceptance and reconstruction efficiency in the lifetime and yield ratio measurements.
The acceptance and efficiency are obtained with an embedding Monte Carlo (MC) technique.
(Anti)hypernucleus decay and the paths of their daughters are simulated using the GEANT3 package~\cite{Brun:1987ma}, taking into account the geometry and materials of the STAR detectors~\cite{Fine:2000qx}.
The responses of the detectors and read-out electronics are also simulated, and the final simulated data are embedded into real data events, which are sampled from different data-taking runs to have a good representation of the whole data set used in the analysis.
The number of MC (anti)hypernuclei embedded is 5\% of the multiplicity of the real-data events.
Then the embedded events are processed through the same reconstruction procedures as real data.
After that, the same track and topological requirements as for the real data are applied to the reconstructed MC (anti)hypernuclei.
The final reconstruction efficiency $\epsilon$ is calculated as the ratio of the number of reconstructed MC (anti)hypernuclei to the number of input MC (anti)hypernuclei.
This efficiency $\epsilon$ includes particle interaction with materials, the detector acceptance, tracking efficiency and selection efficiency.
Since GEANT3 does not properly consider (anti)nucleus absorption by materials, we simulate the $^{3}\hbox{He}$, $^{3}\overline{\hbox{He}}$, $^{4}\hbox{He}$ and $^{4}\overline{\hbox{He}}$ absorption using GEANT4, and further correct their track efficiency from the official STAR simulation. This correction is $<$3\% for nuclei and $<$5\% for antinuclei.
The fraction of (anti)hypernuclei absorbed by the beam pipe (Be) and insulation gas (N$_{2}$) are estimated to be minimal and can be neglected.

The (anti)hypernucleus reconstruction efficiencies as a function of $L/(\beta\gamma)$ are shown in Fig.~\ref{fig:lifetime_efficiency}, which are used to correct the raw yields in different $L/(\beta\gamma)$ intervals before the exponential fits are conducted to extract the lifetimes.

\begin{figure}
    \centering
    \includegraphics[width=0.75\textwidth]{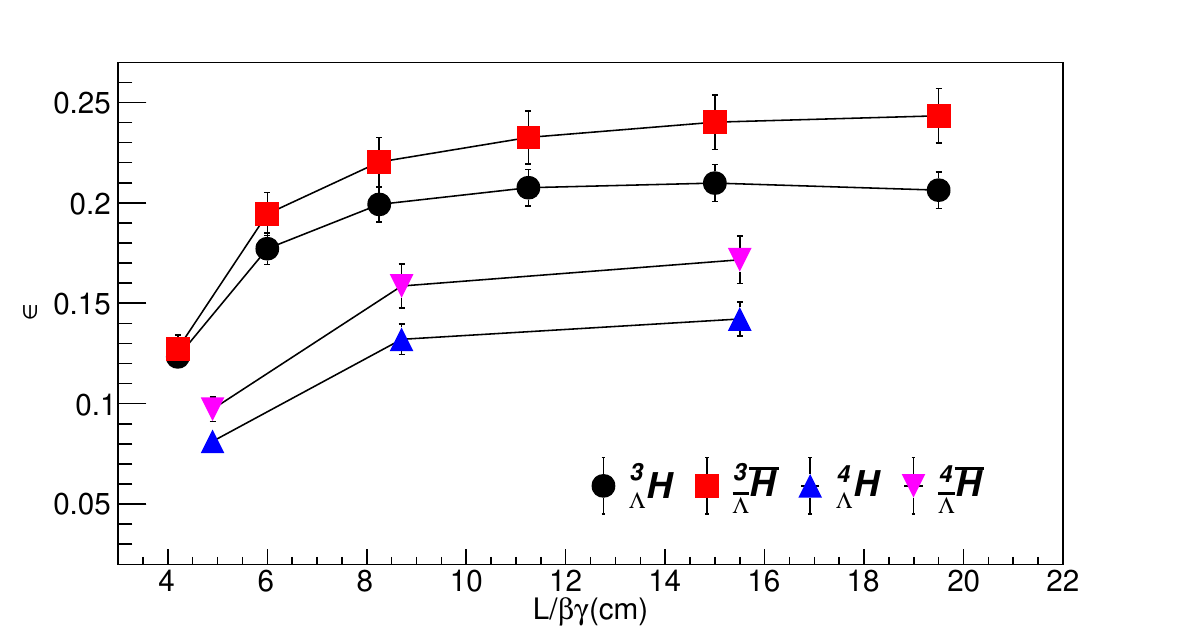}
    \caption{Reconstruction efficiency as a function of $L/(\beta\gamma)$ obtained from the embedding Monte Carlo technique. Hypernuclei have stricter topological cuts than antihypernuclei to suppress knock-out $^3\hbox{He}$ and $^4\hbox{He}$, resulting in lower efficiencies. }
    \label{fig:lifetime_efficiency}
\end{figure}

\subsection{(Anti)hypernuclei, $\Lambda$ and $\Bar{\Lambda}$ lifetime measurements.}
Figure ~\ref{fig:Invmass_tau} shows the invariant-mass distributions of \htl, \ahtl, \hfl~ and \ahfl~ candidates in different $L/(\beta\gamma)$ intervals, which are used to extract their lifetimes. In order to avoid the low transverse momentum region, where the reconstruction efficiency approaches zero and may have relatively large systematic uncertainties, the measurement is performed only for (anti)hypernuclei with $p_{T}>$ 2.1 GeV/$c$. 

\begin{figure}
    \centering
    \includegraphics[width=0.82\textwidth]{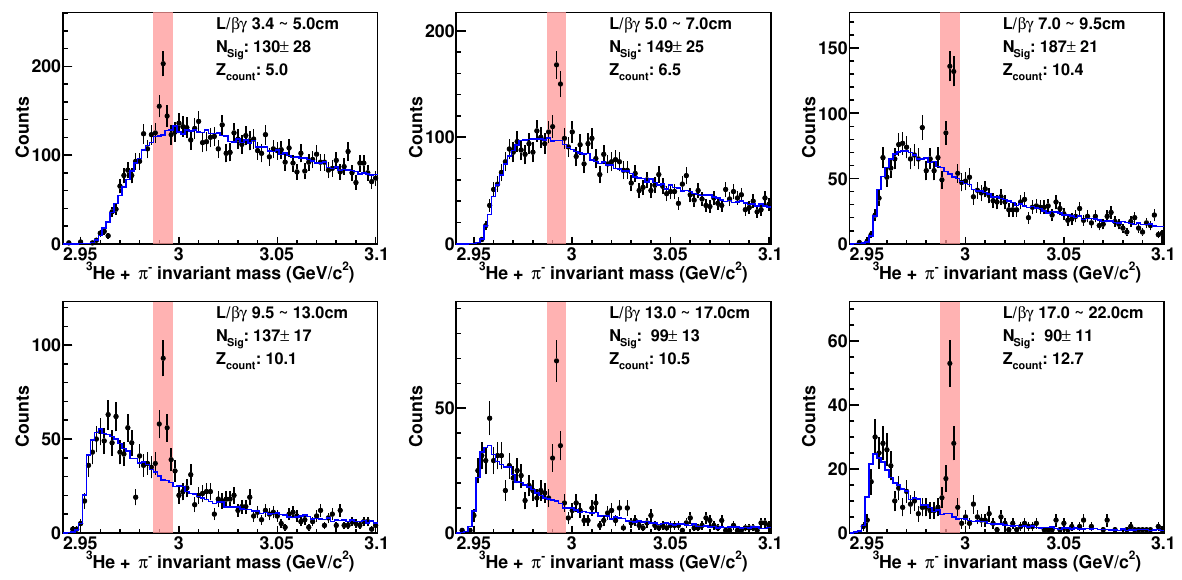}
    \includegraphics[width=0.82\textwidth]{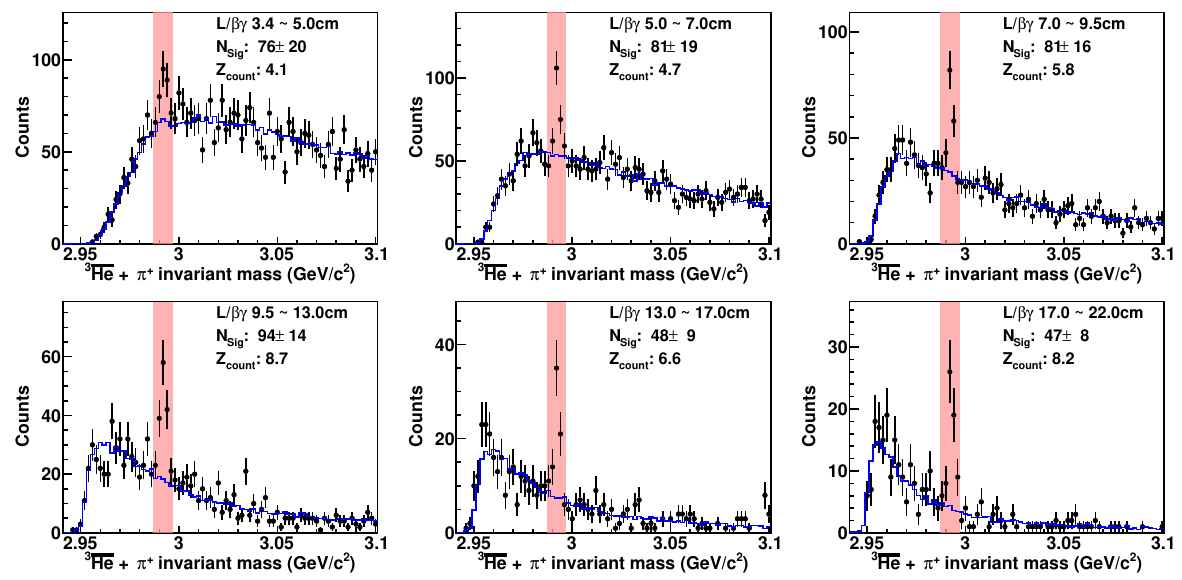}
    \includegraphics[width=0.82\textwidth]{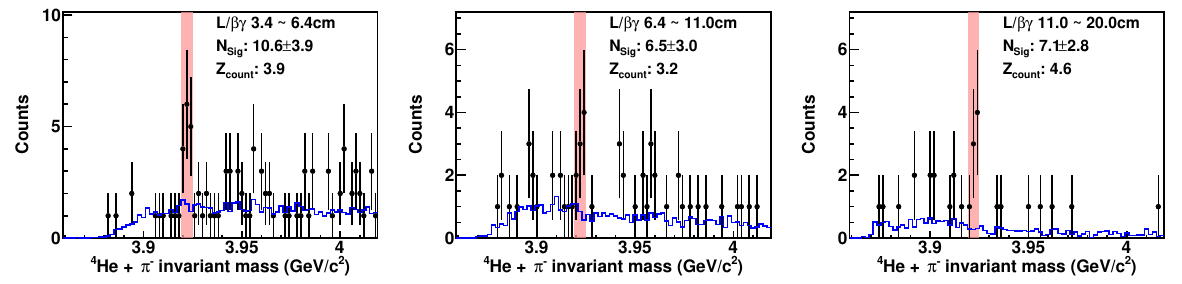}
    \includegraphics[width=0.82\textwidth]{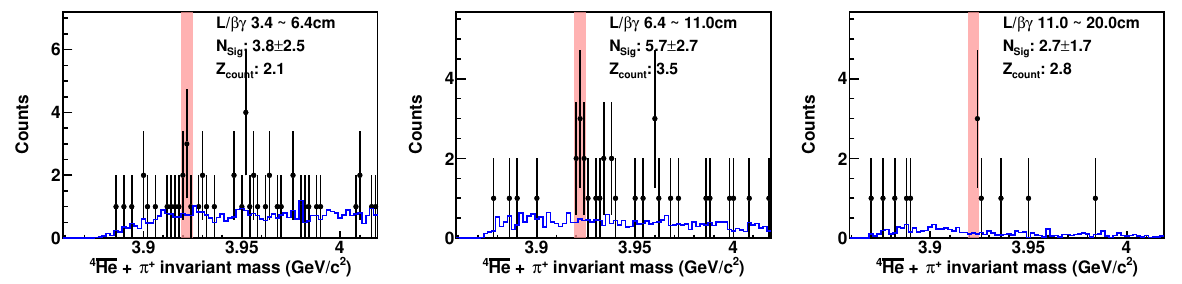}
    \caption{\htl, \ahtl, \hfl~ and \ahfl~ candidate invariant-mass distributions in different $L/\beta\gamma$ intervals.}
    \label{fig:Invmass_tau}
\end{figure}

As an additional test of (anti)hypernucleus lifetime measurements, we have also measured the $\Lambda$ and $\Bar{\Lambda}$ lifetimes with the same method.
3.2 million Au+Au collision events at $\sqrt{s_{NN}}=200$ GeV are used for these measurements.
The topological cuts used to obtain the $\Lambda$ signal are the same as those used in the (anti)hypernucleus analysis, except that an additional $DCA_{V_{0}}<0.1$~cm topological cut is added. $DCA_{V_{0}}$ is the distance-of-closest approach between the reconstructed mother-particle trajectory and the primary vertex. The $DCA_{V_{0}}$ cut suppresses contributions of $\Lambda$ ($\Bar{\Lambda}$) from $\Xi$ ($\Bar{\Xi}$) and $\Omega$ ($\Bar{\Omega}$) decays, which make the measured lifetime longer. This is verified by the fact that the measured $\Lambda$ and $\Bar{\Lambda}$ lifetimes increase as the allowed $DCA_{V_{0}}$ range is enlarged.
Figure~\ref{fig:LambdaLifetime.pdf} shows the $\Lambda$ and $\Bar{\Lambda}$ $L/\beta\gamma$ distributions, and the exponential fits to obtain their lifetimes.
Our measured lifetimes for $\Lambda$ (264.5$\pm$1.6 $ps$) and $\Bar{\Lambda}$ (268.3$\pm$2.3 $ps$) are consistent considering uncertainties, as expected by the $CPT$ symmetry. However, they are slightly longer than the value from the Particle Data Group 263$\pm$2 $ps$~\cite{WorkMan_pdg}. This is expected because the $DCA_{V_{0}}$ cut can not exclude all $\Lambda$ from $\Xi$ and $\Omega$ decays.
No particle yet discovered decays weakly to \htl~or \hfl, so we do not consider the decay feed-down effect for (anti)hypernuclei lifetime measurements in this analysis.

\begin{figure}
    \centering
    \includegraphics[width=0.6\textwidth]{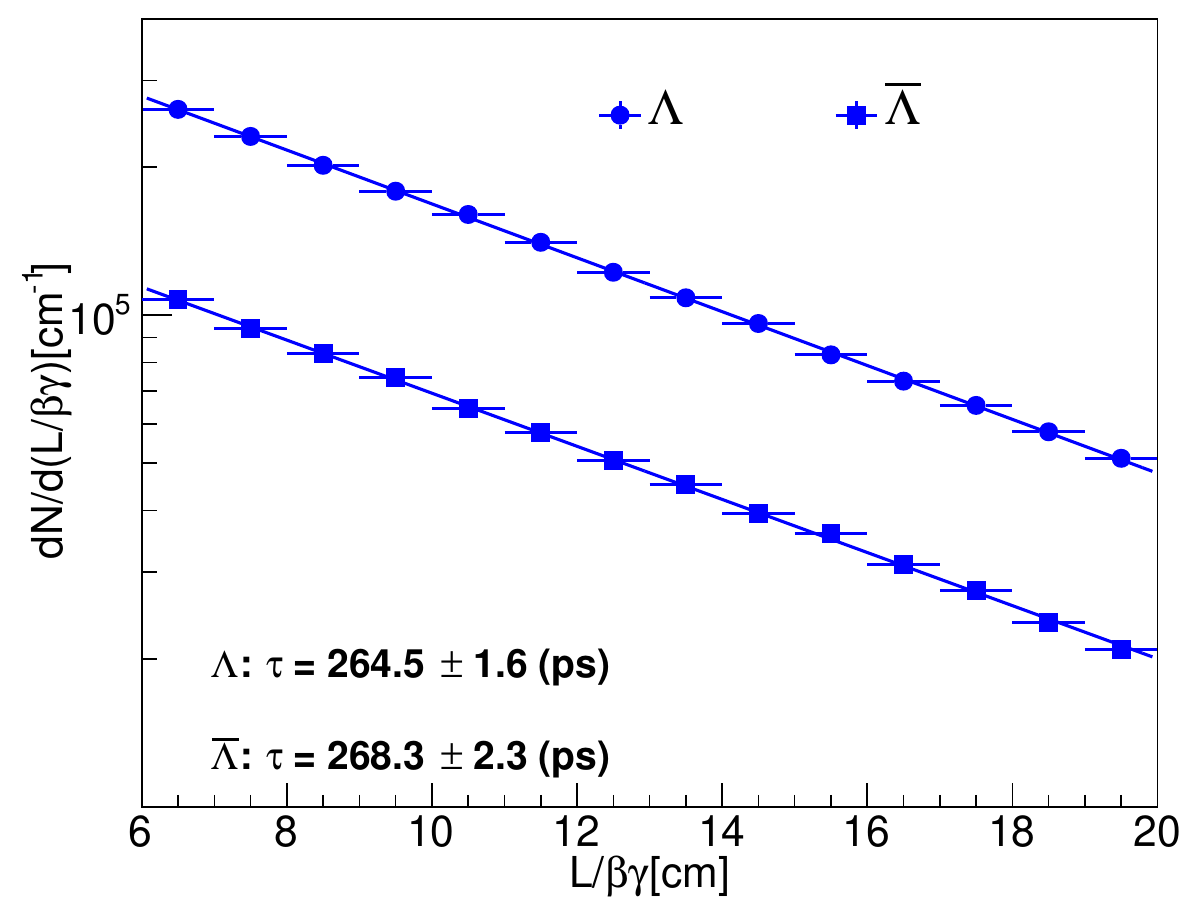}
    \caption{$dN/d(L/\beta\gamma)$ as a function of $L/\beta\gamma$ for $\Lambda$ and $\bar{\Lambda}$, and exponential fits to obtain their lifetimes.}
    \label{fig:LambdaLifetime.pdf}
\end{figure}

\subsection{Yield Measurements.}
The yields of all the studied particles in this work are measured in the phase space of $|y|<0.7$ and $0.7<p_{\rm T}/m<1.5$ with only MB triggered events in order to avoid possible bias from the trigger selection. 
Thus the (anti)hypernucleus signal counts are less than those in Fig. 3 in this paper. The signal and background counts that are used to extract (anti)hypernuclei yield ratios are listed in the Tab.~\ref{counts_yield}.

\begin{table}[htbp]
    \centering
	\caption{The signal and background counts in the measured phase space with MB triggered events.}
	\label{counts_yield}
	\begin{tabular}{cccccc}
		\hline
            Collision systems    &               & \htl          & \ahtl         & \hfl           &\ahfl  \\
		\hline
            Total     &   $N_{Sig}$   & $606\pm42$    & $317\pm31$    & $13.3\pm4.1$   & $8.3\pm3.3$\\
                      &   $N_{Bg}$    & $1145\pm6$    & $605\pm5$    & $3.9\pm0.3$    & $2.7\pm0.3$ \\
            \hline
            Au+Au, U+U   &$N_{Sig}$   &$207\pm27$     & $89\pm19$     &-               &-          \\
                         &$N_{Bg}$    &$517\pm5$     & $267\pm4$      &-               &-          \\
            \hline
            Zr+Zr, Ru+Ru &$N_{Sig}$   &$400\pm32$     &$228\pm24$     &-   &-\\
                         &$N_{Bg}$    &$627\pm4$      &$339\pm3$      &-   &-\\
            \hline
	\end{tabular}
\end{table}

After $|n_{\sigma^{3}\rm{He}}|<3$ and $1<m^{2}/Z^{2}<3$ (GeV/$c^2$)$^2$ selections, 
the $^3\hbox{He}$ and $^3\overline{\hbox{He}}$ candidates are counted with a $1/\epsilon$ weight to get the yield in the measured phase space.

For \htl~and \ahtl~yield measurements, invariant-mass distributions are obtained with a candidate-by-candidate $1/\epsilon$ weight. Then the signal yield is extracted by subtracting the combinatorial background, obtained by the rotation method, from the candidate invariant-mass distribution in the signal range.

For $^4\hbox{He}$, $^4\overline{\hbox{He}}$, \hfl~and \ahfl, the statistics are too low to apply a candidate-by-candidate efficiency correction. We thus calculated the total raw yields in the whole selected $p_{\rm T}$ range and corrected it by the average efficiency. 
The average efficiency is obtained based on knowledge of the $p_{\rm T}$ spectra of $A=3$ (anti)(hyper)nuclei.
Firstly, the $p_{\rm T}$ spectra for $^3\hbox{He}$, $^3\overline{\hbox{He}}$, \htl~and \ahtl~ are obtained and fitted with Blast-Wave~(BW) functions~\cite{Schnedermann:1993ws}
\begin{linenomath*}
\begin{equation}
        \frac{1}{2\pi p_{T}} \frac{d^{2}N}{dp_{T}dy} \propto \int_{0}^{R} r dr m_{0} I_{0} \left(\frac{p_{T} \sinh \rho}{T}\right)K_{1}\left(\frac{m_{T} \cosh \rho}{T}\right),
\end{equation}
\end{linenomath*}
as shown in Fig.~\ref{fig:pTSpectra_all.png}.
Here $\rho=\tanh^{-1}(\beta_{s}(r/R)^{n}$) and $n=1$. The fireball radius $R$ is fixed to 10 fm. $I_{0}$ and $K_{1}$ are Bessel functions. $m_0$ is the particle mass, and $m_{T}=\sqrt{m_{0}^{2}+p_{T}^{2}}$. $\beta_{s}$ and $T$ are free fitting parameters, representing the expansion velocity and temperature of the fireball. 
We then assume the BW functions for $^4\hbox{He}$, $^4\overline{\hbox{He}}$, \hfl~and \ahfl~ have the same $\beta_{s}$ and $T$ as for $^3\hbox{He}$, $^3\overline{\hbox{He}}$, \htl~and \ahtl, respectively, and the only difference in the BW functions are the particle masses~\cite{HypernuclearDataBase}. 
The efficiencies for $^4\hbox{He}$, $^4\overline{\hbox{He}}$, \hfl~and \ahfl~ in the whole measured $p_{T}$ range are calculated as the average efficiency with the above BW-function weights.
The measured raw yields of $^4\hbox{He}$, $^4\overline{\hbox{He}}$, \hfl~and \ahfl~ are then corrected with the average efficiencies to obtain the reported yields.

\begin{figure}
    \centering
    \includegraphics[width=0.8\textwidth]{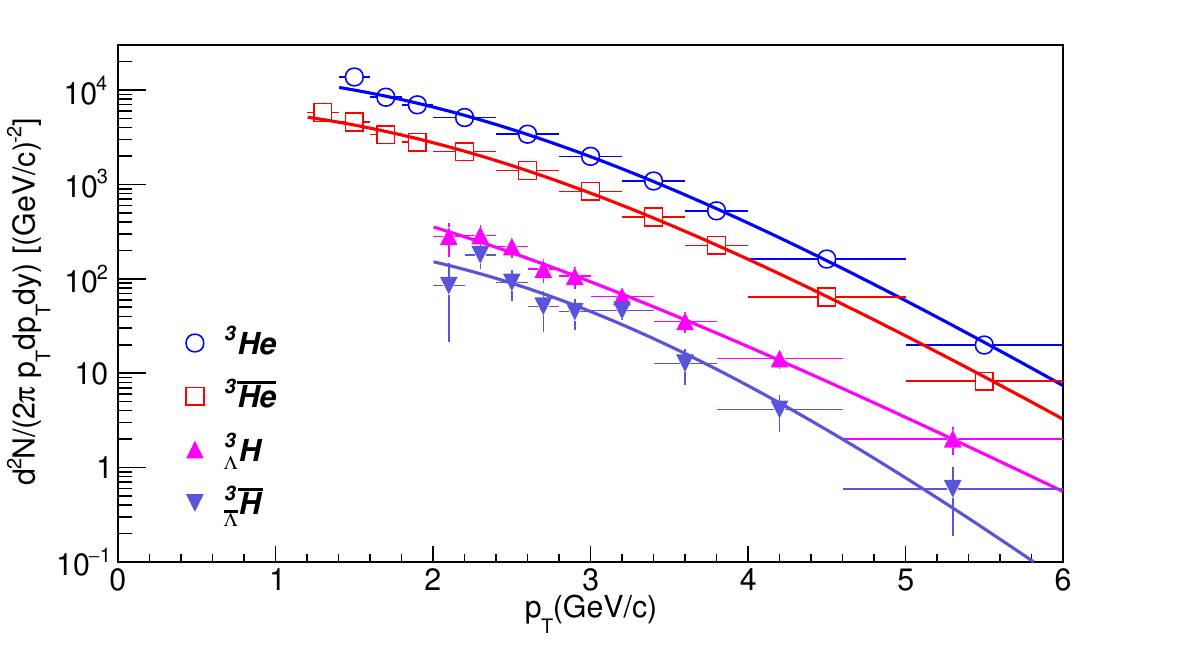}
    \caption{Efficiency corrected $p_{T}$ spectra for $^3\hbox{He}$, $^3\overline{\hbox{He}}$, \htl, and \ahtl. The spectra are in the phase space of $|y|<0.7$ with only MB-triggered events.  The spectra are not normalized by the number of events. The lines represent the BW-function fits.}
    \label{fig:pTSpectra_all.png}
\end{figure}

The yields of $^3\hbox{He}$, $^3\overline{\hbox{He}}$, $^4\hbox{He}$ and $^4\overline{\hbox{He}}$ are also corrected for the contributions from the weak decays of \htl, \ahtl, \hfl~and \ahfl, whose fractions out of the total measured (anti)helium nuclei yields are listed in Tab. ~\ref{fd_percent}.

\begin{table}[htbp]
    \centering
	\caption{Fraction of (anti)helium nuclei from the two-body weak decays of (anti)hypernuclei in different collision systems. The two-body decay branching fractions are assumed to be 0.25 for \htl~ and \ahtl, 0.5 for \hfl~ and \ahfl .}
	\label{fd_percent}
	\begin{tabular}{ccccc}
		\hline
            Collision systems          & $^3{\hbox{He}}$   &$^3\overline{\hbox{He}}$  & $^4{\hbox{He}}$      &$^4\overline{\hbox{He}}$  \\
		\hline
            Total           &(4.3$\pm$0.8)\%    &(4.9$\pm$1.1)\%    &(29$\pm$12)\%   &(42$\pm$21)\%\\
            Au+Au, U+U      &(5.5$\pm$1.7)\%     &(6.2$\pm$2.5)\%     &-      &-\\
            Zr+Zr, Ru+Ru    &(3.6$\pm$1.0)\%      &(4.0$\pm$1.5)\%      &-      &-\\
            \hline
	\end{tabular}
\end{table}

While the measured particle ratios are consistent with previous measurements, we also note that the $^3_{\Lambda}\hbox{H}/^3\hbox{He}$ and $^3_{\bar{\Lambda}}\overline{\hbox{H}}/^3\overline{\hbox{He}}$ ratios in U+U and Au+Au collisions are lower than previous STAR results~\cite{STAR:2010gyg} by 2.8 and 1.9 $\sigma$, respectively.
We have investigated possible sources of the differences. 
The previous analysis used a mixture of MB and  central triggered events. 
The ratios are expected to be higher in central events~\cite{Sun:2018mqq,Bellini:2018epz}. 
The two analyses are also done in slightly different $p_{T}$ ranges. 
These differences alone are not enough to explain  the observed difference between the measured ratios at their face values. 

\subsection{Systematic Uncertainties.}
Four major sources of systematic uncertainties are evaluated for the (anti)hypernucleus-lifetime measurements and the yield-ratio measurements: 
A. Systematic uncertainties on track-reconstruction efficiency, evaluated by varying the minimal number of measured points on the tracks;
B. Systematic uncertainties on (anti)hypernucleus reconstruction efficiency due to topological selections, evaluated by varying the topological-selection viarables;
C. Systematic uncertainties on (anti)hypernucleus signal-yield extraction from the invariant-mass spectra, evaluated by enlarging the invariant-mass ranges for signal-yield integration; and systematic uncertainties from the $p_{T}$-spectrum shapes, evaluated by narrowing the $p_{T}$-spectrum fit ranges;
D. Systematic uncertainties on the (anti)helium yields, evaluated by varying the minimal number of measured points for $\langle{dE/dx}\rangle$ calculation and the cut on the helium-track $DCA$ to primary vertex.
The total systematic uncertainty is calculated as the quadratic sum of the four contributions above.
The systematic uncertainty contributions from different sources for lifetime and yield-ratio measurements are summarized in Tab.~\ref{lifetime_sys}, Tab.~\ref{ratio_sys} and Tab.~\ref{ratio_sub_sys}. When calculating the yield ratios, lifetimes and lifetime differences, the correlations of systematic uncertainties from the same sources have been considered. Thus part of systematic uncertainties will be canceled.

\begin{table}[htbp]
    \centering
	\caption{Systematic uncertainties on (anti)hypernucleus lifetimes.}
	\label{lifetime_sys}
        \begin{tabular}{lllll}
            \hline
            Sources   &$\tau(\HS)$ &$\tau(\aHS)$ &$\tau(\HF)$ &$\tau(\aHF)$ \\
            \hline
            Track reconstruction                &2.8\%  &8.9\%   &15.5\%   &16.8\%\\
            Topological selection               &4.5\%  &7.3\%  &11.9\%   &10.5\%\\
            Signal extraction \& $p_{T}$ shape  &0.4\%  &0.5\%   &2.4\%    &3.8\%\\
            Total                               &5.4\%  &11.6\%  &19.7\%  &20.1\%  \\   
            \hline
	\end{tabular}
\end{table}

\begin{table}[htbp]
    \centering
	\caption{Systematic uncertainties on yield ratios in all measured collision systems.}
	\label{ratio_sys}
	\begin{tabular}{lllllllll}
		\hline
            Sources   & $\frac{^3\overline{\hbox{He}}}{^3{\hbox{He}}}$     &$\frac{^4\overline{\hbox{He}}}{^4{\hbox{He}}}$   &$\frac{\aHS}{\HS}$     
            &$\frac{\aHF}{\HF}$    &$\frac{\HS}{^3{\hbox{He}}}$     &$\frac{\HF}{^4{\hbox{He}}}$     &$\frac{\aHS}{^3\overline{\hbox{He}}}$       &$\frac{\aHF}{^4\overline{\hbox{He}}}$ \\
		\hline
            Track reconstruction                &0.6\%  &0.6\%   &12.6\%   &12.6\%  &5.8\%      &5.8\%  &10.8\%  &10.8\%  \\
            Topological selection               &0.6\%  &0.6\%   &11.4\%   &11.4\%  &3.8\%      &3.8\%  &13.7\% &13.7\% \\
            Signal extraction \& $p_{T}$ shape  &0.1\%  &22.2\%  &1.9\%    &46.3\%  &6.0\%      &20.4\% &8.2\%  &49.9\%\\
            (Anti)helium yields                 &0.3\%  &0.3\%   &-        &-       &3.4\%      &3.4\%  &3.2\%  &3.2\% \\
            Total                               &0.9\%  &22.2\%  &17.1\%   &49.3\%  &9.8\%      &21.8\% &19.5\% &52.9\%\\
            \hline
        \end{tabular}
\end{table}

\begin{table}[htbp]
    \centering
	\caption{Systematic uncertainties on yield ratios in big and small collision systems.}
	\label{ratio_sub_sys}
        \begin{tabular}{lllll}
            \hline
            & \multicolumn{2}{c}{Au+Au, U+U} & \multicolumn{2}{c}{Zr+Zr, Ru+Ru} \\
            Sources   &$\frac{\HS}{^3{\hbox{He}}}$ &$\frac{\aHS}{^3\overline{{\hbox{He}}}}$ &$\frac{\HS}{^3{\hbox{He}}}$ &$\frac{\aHS}{^3\overline{{\hbox{He}}}}$\\
            \hline
            Track reconstruction                &8.1\% &27.0\%         &3.6\%  &4.9\%\\
            Topological selection               &7.0\%  &28.9\%         &3.7\%  &7.9\%\\
            Signal extraction \& $p_{T}$ shape  &15.1\% &18.3\%         &3.0\%  &0.6\%\\
            (Anti)helium yields                 &4.2\%  &3.5\%          &3.8\%  &1.9\%\\
            Total                               &19.0\% &43.7\%         &7.1\% &9.5\% \\
            \hline
	\end{tabular}
\end{table}

\end{methods}

\section*{Reference}
\bibliographystyle{naturemag}